\documentstyle[preprint,eqsecnum,aps]{revtex}
\draft
\preprint{UCT-TP 206/94}
\def\emline#1#2#3#4#5#6{%
       \put(#1,#2){\special{em:moveto}}%
       \put(#4,#5){\special{em:lineto}}}
\begin{document}
\title{Relating the pion decay constant to the
chiral restoration temperature.}
\author{N. Bili\'c\cite{addB},  J. Cleymans and
M.D.~Scadron\cite{Add}}
\address{Physics Department, University of Cape Town,
         Rondebosch, South Africa 7700}
\date{21 January 1994}
\maketitle
\begin{abstract}
We review the
relationship between the pion decay constant $f_\pi$, the chiral
symmetry restoration temperature $T_c$ and the phenomenology
of low energy chiral symmetry breaking in view of the recent
confirmation of the existence of a sigma meson with a mass
of 600-700 MeV.
\end{abstract}

\section{Introduction}
Over the past few years many physicists have appreciated the
importance of a chiral-symmetry restoring phase transition
from nuclear to quark matter.
Since the pion decay constant $f_\pi$ sets the scale for
chiral symmetry breaking, it is natural
to search for a relationship between $f_\pi$ and
the chiral-symmetry restoring temperature $T_c$. The fact that
both quantities are of the same order of magnitude indicates that a
natural relationship between the two could exist.
This relationship was investigated in Refs.~[1,2,3] where it
was proposed that
\begin{equation}
T_{c} = 2f_{\pi}\approx 180 {\rm MeV}
\label{one}
\ .
\end{equation}
for a pion decay constant $f_\pi\approx 90$ MeV in the chiral limit.
This energy scale (1) is compatible with the
estimates from numerical simulations of lattice
gauge theories [4,5].
Yet subsequent studies [6,7,8] of $T_c$ based on the same chiral
four-fermion Nambu - Jona-Lasinio (NJL)
[9] and linear sigma (LSM)
[10] models used in part to derive (1) find
instead $T_c=\sqrt{2}f_\pi$ or
$f_\pi$.

In this paper we return to these two chiral models and demonstrate
that the preferred $T_c$ in (1) is directly correlated to the NJL zero
temperature and chiral-limiting (CL) scalar sigma mass [9]
\begin{equation}
m_\sigma = 2m_q
\label{two}
\ .
\end{equation}
Here $m_q$ is the CL nonstrange (constituent) quark mass near
$m_q\approx M_N/3 \sim 300$ MeV. Since a very recent measurement of
$m_\sigma$ finds [11] a clear enhancement in the invariant
$\pi^0\pi^0$ mass around 650 MeV,
supporting earlier observations [12],
this leaves little doubt that the $\sigma$ (600-700) really
does exist and that the NJL-LSM $\sigma$ meson in (2) is {\bf not}
just a
``toy'' particle.
It is therefore an appropriate time
to return to a study of (1) based on the NJL relation (2).

In Section 2 we review the NJL scheme, primarily focusing on the
relation between $T_c$ and the ultra-violet cutoff $\Lambda$.
Keeping  careful track  of the expansion terms in quark mass
relative to the cutoff $\Lambda$,
one obtains Eq.~(1) after some straightforward algebra.
In Section 3 we work with
a linearized chiral quark model (CQM) lagrangian to obtain the same
results as in the NJL framework. In fact the latter CQM approach can
be used to dynamically generate the complete LSM lagrangian
(including quartic terms) of Gell-Mann and L\'evy. Finally in Section
4 we briefly recall the BCS analogy [1] between energy gap and
critical
temperature. Such an approach is extended to QCD. In all cases in
Sections 2-4 the relation $T_c\simeq 2f_\pi$ is obtained as noted in
the
concluding Section 5.
\bigskip
\section{Four fermion NJL model}
\medskip
We
will start our discussion about the relationship between $T_c$ and
$f_\pi$ in the NJL model. We will show that the correct relationship
is indeed Eq. (1).
To make the point clearly we
adhere in this section to the notation used in Ref.~[7].
The lagrangian density is given by
\begin{equation}
{\cal{L}}_{NJL}
=\bar{\psi}i\not\mkern-4mu{\partial}\psi+
 G\left[
 (\bar{\psi}\psi)^2  +
 (\bar{\psi}i\gamma_5\psi)^2
 \right]\; ,
\label{three}
\end{equation}
where the spinor $\psi$ denotes an isodoublet of u and d
quark fields.
 The standard NJL gap equation for the
nonstrange quark mass depicted in Fig. 1 is given by
\begin{equation}
m_q = \; \tilde{G}
\int {d^4p\over (2\pi)^4} \; {\rm{Tr}}[S_F(p)] \; ,
\label{four}
\end{equation}
where in the notation of Ref.~[7]
$\tilde{G}\equiv G(2N_cN_f+1)$.

This is  generalized to
finite temperatures, using the standard procedure,
i.e., $S_F(p)$ is replaced by the temperature-dependent
propagator
\begin{equation}
S_{F}(p,T) = {i(\not\mkern-4mu{p} + m)\over p^{2} - m^{2}} -
{2\pi \delta (p^{2} - m^{2}) (\not\mkern-4mu{p} + m)\over e^{|p_0/T|}
+ 1}
\; ,
\label{five}
\end{equation}
and the quark mass $m_q$ in (4) is replaced by a temperature-
dependent mass
$m_q(T)$.
As was shown by Dolan and Jackiw [13], this is equivalent to
using the discrete sum over energies in the Matsubara formalism [14].
After these two replacements for $m_q$ and $S_F$ in (4) are made
and the integral over $p_0$ is performed, the
temperature-dependent gap equation
becomes
\begin{equation}
m_q(T) = 4\tilde{G} \int {d^3p\over (2\pi)^3}
{m_q(T)\over 2E_p}\left[1-{2\over e^{E_p/T}+1}\right]
\; ,
\label{six}
\end{equation}
where $E_p\equiv \sqrt{p^2+m_q^2(T)}$ .
A common factor of $m_q(T)$ can be cancelled on both sides of
Eq.~(6). In the limit where one reaches the transition point for
chiral-symmetry restoration, the quark mass ``melts'' and Eq. (6)
becomes
\begin{equation}
1 = 2\tilde{G} \int {d^3p\over (2\pi)^3}
{1\over p}\left[1-{2\over e^{p/T_c}+1}\right]
\; .
\label{seven}
\end{equation}
The integral in (7) can be easily evaluated
using a three-dimensional (3D) cutoff $\Lambda$ , leading to
\begin{equation}
1 = 2\tilde{G}\left[{\Lambda^2\over 4\pi^2}
-{T_c^2\over 12} \right]
\; ,
\label{eight}
\end{equation}
or
\begin{equation}
{T_c^2\pi^2\over 3} = \Lambda^2 - {2\pi^2\over \tilde{G}}
\; .
\label{nine}
\end{equation}
We observe that a linear relation
is established in (9) between the cutoff  $\Lambda^2$
 and the critical temperature $T_c^2$.
For further reference we note that the
integration of the gap equation (4) at zero
temperature gives in terms of this 3D cutoff $\Lambda$
\begin{equation}
1={2\tilde{G}\over 4\pi^2}
\left[
\Lambda\sqrt{{\Lambda^2 + m_q^2}}
-m_q^2
\ln
\left({\Lambda\over m_q}+\sqrt{ 1+{\Lambda^2\over m_q^2}}\right)
\right]
\; .
\label{ten}
\end{equation}

To find the relation between the critical temperature $T_c$, and the
pion decay constant $f_\pi$, we now turn our attention to the
 (logarithmically divergent) ``gap equation''
derived from the quark loop for
$\langle 0|A_{\mu}|\pi \rangle =if_{\pi}q_{\mu}$
in the
chiral limit as
\begin{equation}
f_\pi^2 = -4iN_{c} m_q^2\;
\int {d^4p\over (2\pi)^4}
{1\over (p^2- m_q^2)^2} \; .
\label{eleven}
\end{equation}
Upon integration, (11) leads to
\begin{equation}
f_\pi^2={N_cm_q^2\over 2\pi^2}
\left[
\ln\left( {\Lambda\over m_q}+\sqrt{1+{\Lambda^2\over m_q^2}}\right)
-{\Lambda\over\sqrt{\Lambda^2+ m_q^2}}
\right]
\; .
\label{twelve}
\end{equation}
Combining (12) with the zero-temperature gap
equation (10) one obtains
\begin{equation}
{4\pi^2\over 2\tilde{G}} + {2\pi^2 f_\pi^2\over N_c} =
{\Lambda^3\over \sqrt{\Lambda^2 + m_q^2}}
\;   .
\label{thirteen}
\end{equation}
To proceed further it is necessary to write the
right hand side (rhs) of
(13) in a Taylor series expansion as
\begin{equation}
{\Lambda^3\over \sqrt{\Lambda^2 + m_q^2}}
\approx\Lambda^2 -{1\over 2} m_q^2 +{3\over 8}{m_q^4\over
\Lambda^2}+\cdots
\label{fourteen}
\end{equation}
Now substituting
the cutoff $\Lambda^2$ in (13) into the expression for the
critical temperature in (9) one finds
\begin{equation}
{4\pi^2\over 2\tilde{G}} + {2\pi^2 f_\pi^2\over N_c} =
{T_c^2\pi^2\over 3} +{2\pi^2\over \tilde{G}}-
{1\over 2} m_q^2-{3\over 8}{m_q^4\over \Lambda^2}+\cdots
\; ,
\label{fifteen}
\end{equation}
with terms of
$m_q^6/\Lambda^4$  being neglected .
The last equation can be written in an
almost cutoff-free form
\begin{equation}
T_c^2=2f_\pi^2{3\over N_c}+{3\over 2\pi^2}m_q^2
-{9\over 8\pi^2}{m_q^4\over \Lambda^2}+\cdots
\; ,
\label{sixteen}
\end{equation}
a result which was first obtained by Koci\'c [15]
(to leading order).

Upon using the quark-level
Goldberger-Treiman (GT) relation
\begin{mathletters}
\begin{equation}
m_q=f_\pi g
\;  ,
\label{seventeena}
\end{equation}
along with
the dimensionless pion-quark coupling constant [16]
\begin{equation}
g={2\pi\over\sqrt{N_c}}
\;  ,
\label{seventeenb}
\end{equation}
\end{mathletters}
Eq.~(16) becomes (for $N_c=3$)
\begin{equation}
T_c^2=2f_\pi^2+2f_\pi^2-{9\over 8\pi^2}{m_q^4\over \Lambda^2}+\cdots
\; .
\label{eighteen}
\end{equation}
Finally dropping the small ${\cal{O}}(m_q^4/\Lambda^2)$ term,
Eq. (18) gives the critical temperature
\begin{equation}
T_c\simeq 2f_\pi\approx 180 {\rm MeV}
\;  .
\label{nineteen}
\end{equation}
Keeping the last term in Eq. (18) reduces $T_c$
from 180 MeV to 172 MeV.  Although (19) was originally determined in
an NJL scheme in Ref.~[2], it also follows
in the recent review [7] if all
leading order terms are consistently kept.
\bigskip
\section{NJL - LSM model}
\medskip
The above NJL analysis scaled to a 3-dimensional cutoff can in fact
be formulated in the equivalent language of linear $\sigma$ and $\pi$
auxiliary (pseudo-elementary)
fields with the following  lagrangian density exploited by Eguchi [17]
\begin{equation}
{\cal{L}}_E
=\bar{\psi}i\not\mkern-4mu{\partial}\psi
+ g \bar{\psi}(\sigma + i \gamma_{5}
               \vec{\tau}\cdot\vec{\pi}) \psi
  -{\mu_\sigma^2\over 2}\left[ \sigma ^2+ \pi^2\right]
\; .
\label{twenty}
\end{equation}
Using path integral methods it has been shown that the lagrangians
(3) and (20) are identical provided
the dimensional coupling G of the NJL model is related to
the dimensionless meson-quark coupling $g$ as
$
G=g^2/2\mu_\sigma^2
$.
With hindsight , the crucial relation used here (17b), may appear to
be outside the framework of the NJL model. But its relation to the
NJL coupling $G$ suggests
\begin{equation}
\tilde{G}\approx N_c/2f_\pi^2
\; .
\label{twentyone}
\end{equation}
This means that $\tilde{G}$ scaling the mass gap equation (4), is
itself scaled to (21). Furthermore since
$f_\pi\sim N_c^{1/2}$, $\tilde{G}$ is independent of $N_c$, as
expected.

Recently the equivalence
between (3) and (20) has been used by Delbourgo and Scadron [18]
to combine the features of the NJL model with
some of those of the linear $\sigma$  model.
In this chiral quark model (CQM) context the lagrangian density
has the chiral-invariant form closely related to (20)
\begin{equation}
{\cal{L}}_{CQM}
=\bar{\psi}i\not\mkern-4mu{\partial}\psi
  +{1\over 2}\left[ (\partial\sigma )^2+(\partial\pi )^2\right]
%  -{\mu^2\over 2}\left[ \sigma ^2+ \pi^2\right]
+ g \bar{\psi}(\sigma + i \gamma_{5}
               \vec{\tau}\cdot\vec{\pi}) \psi
\; .
\label{twentytwo}
\end{equation}
In addition one requires the validity of
a mass  gap equation simulating the
NJL Eq. (4) but also including Eq. (11).
So we consider quark mass generation in the dynamically generated
CQM-LSM framework. Following Refs.~[18,19], the log-divergent ``gap
equation'' derived from the quark loop for
$\langle0|A_{\mu}|\pi \rangle=i~f_{\pi}~q_{\mu}$
 for decay constant
$f_{\pi} =m_{q}/g$
in the CL is (defining
$\bar{d}^{4}p = (2\pi)^{-4} \; d^{4}p$)
\begin{mathletters}
\begin{equation}
1 = -i \; 4 \; N_{c} \; g^{2} \int \; \bar{d}^{4}p \; (p^{2}
- m_{q}^{2})^{-2}
\; .
\label{twentythreea}
\end{equation}

The above nonperturbative equation for $\delta f_{\pi} = f_{\pi}$
should be correlated with $\delta m_{q} = m_{q}$ obtained from the
quadratically divergent [18] quark ``tadpole'' graph of Fig. 2 for
$u$
and $d$ quark flavours and colour number $N_{c} = 3$, leading to the
``mass gap''
\begin{equation}
m_{q} = {i \; 8 \; N_{c} \; g^{2}\over - m_{\sigma}^{2}} \int
{\bar{d}^{4}p \; m_{q}\over p^{2} - m_{q}^{2}}
\; .
\label{twentythreeb}
\end{equation}
\end{mathletters}
The NJL mass gap (4) and the CQM mass gap (23b) are consistent with
the NJL coupling (21).

Note that $m_{q}$ in (23b) is a counter-term mass in the kinetic part
of the chiral symmetric CQM lagrangian
$i{\not\mkern-4mu{\partial}} \rightarrow
i{\not\mkern-4mu{\partial}} - m_{q}$ + ``$m_{q}$''.
In this self-consistent
approach the first $-m_{q}$ term signals the quark loop in
Fig. 2 to have mass $m_{q}$ while the second + ``$m_{q}$''
counter-term mass on the lhs of (23b) is computed from the
quark loop, itself on the rhs of (23b).
Finally, one uses the dimensional regularization identity in $2\ell=4$
dimensions [18]
\begin{equation}
\int {\bar{d}^{4}p\over p^{2} - m^{2}} - \int {\bar{d}^{4}p \;
m^{2}\over
(p^{2} - m^{2})^{2}} = \lim_{\ell \rightarrow 2}
{-i \; m^{2\ell -2}\over (4\pi)^{\ell}} \left[ \Gamma (1 - \ell) +
\Gamma (2 - \ell) \right] = {i \; m^{2}\over (4\pi)^{2}}
\; ,
\label{twentyfour}
\end{equation}
together with the log-divergent gap equation (23a). Then the mass gap
(23b) becomes replaced by
\begin{equation}
1 = {2 \; m_{q}^{2}\over m_{\sigma}^{2}} \left[ 1 + {g^{2} \;
N_{c}\over
4\pi^{2}} \right]
\; .
\label{twentyfive}
\end{equation}
Mass counter-terms coupled with the dimensional regularization
identity (24) lead to cancelling signs which then scales mass
in the CQM in a
manner analogous to mass generation in the NJL scheme. Only now 4D
cutoffs can be explicitly avoided, whereas in the NJL analysis in
Section 2, the use of 3D cutoffs enters the theory.

In a similar manner to the quark loop in Fig.~2, the scalar
(counter-term) mass $m_{\sigma}^{2}$ is self-consistently determined
[18] by the $\sigma$ quark ``bubble'' and tadpole graphs of Fig.~3
leading to the CL value
\begin{mathletters}
\begin{eqnarray}
m_{\sigma}^{2} &=& 8i N_{c} \; g^{2} \; \int \; {\bar{d}^{4}p \;
(p^{2} + m_{q}^{2})\over (p^{2} - m_{q}^{2})^{2}}
- 3! \; {8 \; N_{c} \; g \; g' \; m_{q}\over m_{\sigma}^{2}}
\int \; {\bar{d}^{4}p\over p^{2} - m_{q}^{2}}\\
&=& 16 \; i \; N_{c} \; g^{2} \; \left[ \int {\bar{d}^{4}p \;
m_{q}^{2}\over
(p^{2} - m_{q}^{2})^{2}} - \int \; {\bar{d}^{4}p\over
p^{2} - m_{q}^{2}} \right]  \\
&=& {N_{c} \; g^{2} \; m_{q}^{2}\over \pi^{2}} \ .
\end{eqnarray}
\end{mathletters}
To proceed from (26a) to (26b) we have applied the meson-meson
coupling
$g' = m_{\sigma}^{2}/2f_{\pi}$
needed to keep $m_\pi=0$
in the LSM [10].
Then (26c) follows from (26b) using the dimensional
regularization identity (24).
Finally the two equations (25) and (26c)
can be solved for the two unknowns (for $N_{c} = 3$) as [18]
\begin{mathletters}
\begin{equation}
m_{\sigma} = 2 m_{q}
\label{27a}
\end{equation}
\begin{equation}
g = 2\pi/\sqrt{3} \approx 3.6276
\; .
\label{27b}
\end{equation}
\end{mathletters}
Since the quark tadpole graph of Fig. 2 is the obvious extension of
the NJL four-quark Hartree loop
of Fig. 1, it should not be surprising that both the
dynamically generated NJL and LSM-type theories lead to (27a).

{}From the perspective of the CQM lagrangian (22), the meson-quark
coupling $g$ is also self-consistently determined in (27b). In fact
this coupling $g\approx 3.63$ in (27b) makes direct
contact with the empirically deduced [20] $\pi N N$
pseudoscalar coupling $g_{\pi N N} \approx$ 13.08,  since the latter
implies
\begin{equation}
g = g_{\pi NN}/3g_{A} \approx 3.55
\label{28}
\end{equation}
for 3 quarks in a nucleon,
with [21] $g_{A} \approx$ 1.2573 a three-body effect.
But from our point of view, the most important consequence of (27b)
is that this CQM coupling constant $g$ correctly sets the scale of
the CL quark and sigma masses via $f_{\pi} \approx$ 90 MeV as [18]
\begin{mathletters}
\begin{equation}
m_{q} = (2\pi/\sqrt{3}) f_{\pi} \approx 325
\; {\rm MeV}
\label{29a}
\end{equation}
\begin{equation}
m_{\sigma} = 2m_{q} = (4\pi/\sqrt{3}) f_{\pi} \approx 650 \;
{\rm MeV}
\; .
\label{29b}
\end{equation}
\end{mathletters}
Not only does (29b) comply with recent data [11] on
$m_{\sigma}$, but
(29a) is close to the usual quark model value $M_{N}$/3 and also near
the
gauge-parameter independent nonperturbative QCD dynamical value of
[16]
$m_{dyn} \approx [4\pi \alpha_{s} \langle -\bar{q} q
\rangle/3]^{1\over
3} \approx$ 320 MeV
for the usual quark condensate
$\langle -\bar{q} q \rangle^{1\over 3} \approx$ 250 MeV
at momentum scale 1 GeV where [22] $\alpha_{s}$ (1 GeV$^{2}$)
$\approx$ 0.5.

As for the original chiral-invariant linear sigma model (LSM) of
Gell-Mann and L\'evy [10], starting from the CQM lagrangian (22) the
shifted interacting part of the LSM lagrangian can be
self-consistently induced at tree level as
\begin{mathletters}
\begin{equation}
{\cal{L}}_{int}=
 g \bar{\psi}(\sigma + i \gamma_{5}
               \vec{\tau}\cdot\vec{\pi}) \psi
+
g' \sigma (\sigma^2 + \vec{\pi}^{2})
-(\lambda/4) (\sigma^2 + \vec{\pi}^{2})^2
\; ,
\label{thirtya}
\end{equation}
with chiral couplings
\begin{equation}
g=m_q/f_\pi\; \; ,\;\; g'=\lambda f_\pi = m_\sigma^2/2f_\pi
\; .
\label{thirtyb}
\end{equation}
\end{mathletters}
The first meson-quark coupling term in (30) is self-consistently
determined from the CQM lagrangian (22). However the mesonic cubic
and quartic LSM couplings in (30) obtained in tree order are again
induced in one-loop order via the log-divergent gap equation (23a) as
[18]
\begin{mathletters}
\begin{eqnarray}
g_{\sigma \pi \pi} &=& - i \; 2 \; N_{c} \; g^{3} \; \int
{\bar{d}^{4}p \; {\rm Tr}(\not\mkern-4mu{p} + m_{q})(p^{2} -
m_{q}^{2})\over
(p^{2}-m_{q}^{2})^{3}}
\nonumber \\
&=& 2m_{q} \; g \left[ - i \; 4 \; N_{c} \; g^{2} \int
\bar{d}^{4}p \; [p^{2}-m_{q}^{2}]^{-2}\right]
\nonumber \\
&=& 2m_{q}g \; = g'
\; ,
\label{31a}
\end{eqnarray}
\begin{equation}
\lambda = 2 \; g^{2} \left[ - i \; 4 \; N_{c} \; g^{2} \int
\bar{d}^{4}p \; [p^{2} - m_{q}^{2}]^{-2} \right] = 2g^{2} \;
\; ,
\label{31b}
\end{equation}
\end{mathletters}
provided $f_\pi g = m_q$ and $m_\sigma = 2m_q$.
In effect equations (31) correspond to the ``compositeness
condition'' [23] $Z=0$, treating
the $\pi$ and $\sigma$ states  as bound (in the NJL model) or
as elementary (in the CQM or LSM theories).

The authors of Ref.~[8] considered a
similar starting point, based on
the linear $\sigma$ model. The gap equation (23a)
is obtained via the condition that $Z_3=0$
(also see Ref.~[24]), implicitly leading to (31a).
However imposing an additional compositeness condition
$Z_4=0$, Ref.~[8] do not obtain
(31b) .
Instead they find
 $\lambda = g^2$, $T_c\simeq f_\pi$
along
with $m_\sigma^2=2m_q^2+m_\pi^2$.
All of these results are  in disagreement with our conclusions
above.
In fact their latter result is not consistent with the standard form
$m_\sigma^2 = 4m_q^2 + m_\pi^2$ and this suggests as noted
by the authors of  Ref.~[8]
that a complete LSM has a $\lambda \sigma^4$ term with meson
loops which may not be compatible with a $Z_4=0$ compositeness
condition.

Another approach to the strict LSM based on the Lee null tadpole
condition [25] was considered in Ref.~[6]. Although this analysis
also finds $T_c\approx f_\pi$, the authors of Ref.~[6] also note that
this result is ``physically not relevant in the high-temperature
region''.

As for finite temperature effects in the CQM-LSM framework,
at zero chemical potential
the chiral symmetry restoration temperature $T_{c}$
should be the same whether the quark mass $m_{q}$ or the sigma mass
$m_{\sigma}$ ``melts'' to zero.
We limit our considerations to the melting
of  the quark tadpole mass of Fig.~2, resulting in
\begin{equation}
m_{q} \rightarrow m_{q} \; (T_{c}) = m_{q} +
{8 \; N_{c} \; g^{2} \; m_{q}\over - m_{\sigma}^{2}} \;
\left( {T_{c}^{2}\over 2\pi^{2}} \right) \; J_{+} (0) \; ,
\label{thirtytwo}
\end{equation}
where $J_+(0)$ is one of the exponential integrals
\begin{equation}
J_{+}(0) = \int_{0}^{\infty} \; x \; dx \;(e^x+1)^{-1} =
\pi^{2}/12\; ,
\label{thirtythree}
\end{equation}
so that
($T_{c}^{2}/2\pi^{2}$) $J_{+}(0) = T_{c}^{2}/24$ for fermion
loops. The minus sign in the denominator of (32) is due to the
($q^{2} - m_{\sigma}^{2}$)$^{-1}$ propagator structure of the tadpole
at $q^{2} = 0$ and the two minus signs
from the fermion loop and from the antiperiodicity condition for
fermions cancel for fermion loops.
Since
the quark mass melts at
$m_{q}$~($T_{c}$)=0, and
the $T=0$ $m_{q}$ factor divides out of (32), one then obtains
\begin{equation}
m_{\sigma}^{2} = g^{2} \; T_{c}^{2} \; \; \; \; {\rm or} \; \; \; \;
T_{c} = 2f_{\pi} \; ,
\label{thirtyfour}
\end{equation}
for $N_f=2$, $N_c=3$,
along with $m_{\sigma} = 2 m_{q} = 2f_{\pi} g$.

Alternatively in a spontaneously generated LSM framework, a $\sigma$
mass not constrained to $2m_q$ melts to zero to give once again [1]
$T_c=2f_\pi$. This also follows from the LSM analysis [2] for
$m_\sigma^2(T_c)=0$
if the quark tadpole of Fig. (2) is kept.
We believe it significant that whether melting the quark mass as in
(32)
or melting the $\sigma$ mass as in Refs.~[1] or [2], $T_{c}$ is always
$2f_{\pi}\approx 180$ MeV for the CL pion decay constant $f_{\pi}
\approx$
90 MeV.
\bigskip
\section{BCS Approach}
\medskip
Here we offer one final justification of $T_c=2f_\pi$
based on the original BCS approach to low-temperature
superconductivity [26]. As noted in Refs.~[1,2], BCS cut off
a non-relativistic gap integral
\begin{equation}
1 = \lambda_{D} \int^{\omega_{D}} \; {d^3p\over (2\pi )^32E} \tanh \;
{E\over 2T}
\label{thirtyfive}
\end{equation}
at the Debye energy $\omega_{D} >> T_{c}$. Then BCS made an asymptotic
expansion in the the energy gap
($\Delta_0$) relative to $T_{c}$ to obtain the dimensionless ratio
\begin{mathletters}
\begin{equation}
2 \Delta_0/T_{c} = 2\pi \; e^{- \gamma_{E}} \approx 3.52
\; .
\label{36a}
\end{equation}
But this BCS ratio (36a) is strikingly similar to the relativistic
Goldberger-Treiman ratio at the quark level for the NJL-LSM
\begin{equation}
m_{q}/f_{\pi} = g = 2\pi/\sqrt{3} \approx 3.63
\; ,
\label{36b}
\end{equation}
\end{mathletters}
provided one replaces $\Delta_0$ in (36a) by $m_{q}$ in (36b) and
{\bf identifies $T_{c}$ with 2$f_{\pi}$} (or $T_{c} \approx$ 180 MeV).

Also, an analogue (but unmeasured) Debye high energy cutoff
$\omega_{D}$
for a constant $m_{q}$ numerically requires [1]
$T_{c} \approx$ 176 MeV.
However, a ``running mass'' QCD modification eliminates the need for
this
$\omega_{D}$ cutoff but still results [1] in a (melting) chiral
temperature
$T_{c} \approx$ 170 MeV via a gap-type equation similar to (35)
(using $E_{k}(T) = [\underline{k}^{2} + m_{dyn}^{2}(T)]^{1\over 2}$)
$$
1 = {2\alpha_{s}\over \pi} \left[ \int_{0}^{m_{dyn}}
{dk\over E_{k}(T_{c})} \tanh {E_{k}(T_{c})\over 2T_{c}}
+ \int_{m_{dyn}}^{\infty}
{dk\over E_{k}(T_{c})} {m_{dyn}^{2}\over \underline{k}^{2}}
\tanh {E_k(T_c)\over 2T_c} \right.
$$
\begin{equation}
\left. + \int_{m_{dyn}}^{\infty}
{dk\over E_{k}(T_{c})} \left( {m_{dyn}\over 2|\underline{k}| -
m_{dyn}}
- {m_{dyn}^{2}\over \underline{k}^{2}} \right) \right]
\; ,
\label{37}
\end{equation}
for the dynamically generated QCD quark mass [26] $m_{dyn} \approx$
320 MeV.

The important point to note is that  a nonrelativisitc BCS asymptotic
expansion (36a) involves Cooper-paired acoustical phonons with
$E\sim p$.
Likewise a tightly bound
$\bar{q}q$ pion couples to quarks
in (36b) via  relativistic gluon exchange
with $E\sim p$. Consequently it should not be surprising that the
energy gap to critical temperature ratios in (36) strongly suggest
$T_c=2f_\pi$ in the relativistic case (36b).
\section{Conclusions}
In conclusion, we have reviewed different proposals made recently in
the literature about the relationship between two basic quantities
appearing in theories of chiral symmetry , namely the pion decay
constant $f_\pi$ and the chiral symmetry restoration temperature
$T_c$.  The result which best describes the
analysis of lattice gauge
theories is $T_c=2f_\pi$ and we have discussed various ways in
Sections 2-4  to
justify this relationship. The recent confirmation of the
$\sigma$-meson of a mass approximately 650 MeV is a very important
road indicator towards the correct description of low energy
phenomena using models based on chiral symmetry.
\acknowledgments
The authors are grateful for discussions with D. Luri\'{e}.
One of us (MDS) appreciates the hospitality of the University of Cape
Town
and is grateful for partial support from the US Department of Energy.

\begin{figure}
\caption{Quark mass generation in Hartree approximation in the NJL
model.}
\label{fig1}
\end{figure}
\begin{figure}
\caption{Quark tadpole diagram representing quark mass $m_{q}$
in the CQM-LSM.}
\label{fig2}
\end{figure}
\begin{figure}
\caption{Quark bubble (a) and tadpole diagram (b)
              representing squared $\sigma$ mass $m_{\sigma}^{2}$. }
\label{fig3}
\end{figure}
\vfill
\eject
\unitlength=1.00mm
\special{em:linewidth 0.4pt}
\linethickness{0.4pt}
\begin{picture}(120.00,115.00)
\emline{20.00}{100.00}{1}{60.00}{100.00}{2}
\emline{81.00}{99.00}{3}{120.00}{99.00}{4}
\put(103.00,105.00){\circle{12.17}}
\emline{68.00}{101.00}{5}{72.00}{101.00}{6}
\emline{68.00}{99.00}{7}{72.00}{99.00}{8}
\emline{38.00}{102.00}{9}{42.00}{98.00}{10}
\emline{42.00}{102.00}{11}{38.00}{98.00}{12}
\emline{20.00}{20.00}{13}{60.00}{20.00}{14}
\emline{38.00}{23.00}{15}{42.00}{16.00}{16}
\emline{42.00}{23.00}{17}{38.00}{16.00}{18}
\emline{68.00}{22.00}{19}{72.00}{22.00}{20}
\emline{68.00}{20.00}{21}{72.00}{20.00}{22}
\emline{81.00}{20.00}{23}{120.00}{20.00}{24}
\emline{103.00}{20.00}{25}{103.00}{23.00}{26}
\emline{103.00}{25.00}{27}{103.00}{28.00}{28}
\emline{103.00}{30.00}{29}{103.00}{33.00}{30}
\emline{103.00}{35.00}{31}{103.00}{38.00}{32}
\put(103.00,45.00){\circle{14.00}}
\put(101.00,56.00){\makebox(0,0)[cc]{u , d}}
\put(67.00,69.00){\makebox(0,0)[cc]{Figure 1}}
\put(67.00,0.00){\makebox(0,0)[cc]{Figure 2}}
\put(98.00,30.00){\makebox(0,0)[cc]{$\sigma$}}
\put(103.00,115.00){\makebox(0,0)[cc]{u , d}}
\end{picture}
\vfill
\eject
\begin{picture}(135.00,127.00)
\emline{20.00}{90.00}{1}{23.00}{90.00}{2}
\emline{25.00}{90.00}{3}{28.00}{90.00}{4}
\emline{30.00}{90.00}{5}{33.00}{90.00}{6}
\put(43.00,90.00){\circle{14.00}}
\emline{52.00}{90.00}{7}{55.00}{90.00}{8}
\emline{57.00}{90.00}{9}{60.00}{90.00}{10}
\emline{62.00}{90.00}{11}{65.00}{90.00}{12}
\emline{80.00}{90.00}{13}{84.00}{90.00}{14}
\emline{82.00}{88.00}{15}{82.00}{92.00}{16}
\emline{95.00}{90.00}{17}{98.00}{90.00}{18}
\emline{100.00}{90.00}{19}{103.00}{90.00}{20}
\emline{105.00}{90.00}{21}{108.00}{90.00}{22}
\emline{110.00}{90.00}{23}{113.00}{90.00}{24}
\emline{113.00}{90.00}{25}{115.00}{90.00}{26}
\emline{117.00}{90.00}{27}{120.00}{90.00}{28}
\emline{122.00}{90.00}{29}{125.00}{90.00}{30}
\emline{127.00}{90.00}{31}{130.00}{90.00}{32}
\put(132.00,90.00){\rule{3.00\unitlength}{0.00\unitlength}}
\emline{113.00}{90.00}{33}{113.00}{93.00}{34}
\emline{113.00}{95.00}{35}{113.00}{98.00}{36}
\emline{113.00}{100.00}{37}{113.00}{103.00}{38}
\emline{113.00}{105.00}{39}{113.00}{108.00}{40}
\put(113.00,116.00){\circle{14.00}}
\put(40.00,101.00){\makebox(0,0)[cc]{u , d}}
\put(111.00,127.00){\makebox(0,0)[cc]{u , d}}
\put(24.00,85.00){\makebox(0,0)[cc]{$\sigma$}}
\put(58.00,86.00){\makebox(0,0)[cc]{$\sigma$}}
\put(102.00,87.00){\makebox(0,0)[cc]{$\sigma$}}
\put(127.00,88.00){\makebox(0,0)[cc]{$\sigma$}}
\put(118.00,100.00){\makebox(0,0)[cc]{$\sigma$}}
\put(74.00,35.00){\makebox(0,0)[cc]{Figure 3}}
\put(43.00,69.00){\makebox(0,0)[cc]{(a)}}
\put(115.00,69.00){\makebox(0,0)[cc]{(b)}}
\end{picture}
\end{document}